\documentclass[final,3p,times,twocolumn]{elsarticle}
\biboptions{comma,sort&compress}

\usepackage[utf8]{inputenc}
\usepackage[T1]{fontenc}
\usepackage{lmodern}
\usepackage[english]{babel}
\usepackage{textcomp}

\usepackage{graphicx,color}
\usepackage{amsmath,amsfonts,amssymb}
\usepackage{slashed,stackrel}

\usepackage{feynmp-auto}
\def\fmfstyle{
  \fmfset{thin}{0.3pt}         % 1pt
  \fmfset{thick}{1.5thin}      % 1.5thin
  \fmfset{arrow_ang}{15}       % 15
  \fmfset{arrow_len}{2.0mm}    % 4mm
  \fmfset{curly_len}{1.5mm}    % 3mm
  \fmfset{dash_len}{2.5mm}     % 3mm
  \fmfset{dot_len}{1.0mm}      % 2mm
  \fmfset{dot_size}{1.0mm}     % 4thick
  \fmfset{zigzag_len}{2.0mm}   % 2mm
  \fmfset{zigzag_width}{0.2mm} % 2thick
}

\newcommand{\comment}[1]{}
\newcommand{\order}[1]{\ensuremath{\mathcal{O} \! \left( #1 \right)}}
\renewcommand{\d}[1][]{\ensuremath{\operatorname{d}\!{#1}}}
\renewcommand{\Im}{\operatorname{Im}}
\newcommand{\ep}{\ensuremath{\epsilon}}
\newcommand{\HPL}[2][x]{\operatorname{H}_{#2}\!\left( #1 \right)}

\begin{document}

\begin{frontmatter}

\title{
  \vskip-3cm
  {\baselineskip14pt
    \begin{flushright}
      \normalsize DESY 15-165
    \end{flushright}}
  \vskip1.5cm
  Methods for Higgs boson production at N$^3$LO$^*$}
% \corref{cor0}

\cortext[cor0]{%
  Talk given at 18th International Conference in Quantum Chromodynamics
  (QCD 15, 30th anniversary), 29 June - 3 July 2015, Montpellier - FR}

\author[label1]{%
  J.~Hoff\fnref{fn1}}

\fntext[fn1]{%
  Speaker, Corresponding author.}

\ead{%
  jens.hoff@desy.de}

\address[label1]{%
  Deutsches Elektronen Synchrotron DESY, Platanenallee 6, 15738 Zeuthen,
  Germany}

\pagestyle{myheadings}
% \markright{}
\begin{abstract}
  We discuss methods for the calculation of the total partonic Higgs
  boson production cross section via gluon fusion and the result for the
  contribution that stems from two quarks of different flavor in the
  initial state.  Our calculation is exact in the Higgs boson mass and
  the partonic center-of-mass energy.  The result is expressed in terms
  of iterated integrals, some of which are Harmonic Polylogarithms,
  whereas others are new iterated integrals.  We comment on the methods
  relevant for the reduction to scalar integrals and new types of
  function that appear in the final result.
\end{abstract}


\end{frontmatter}

% -- main text ---------------------------------------------------------

\section{Introduction}

After the Higgs boson discovery~\cite{Aad:2012tfa,Chatrchyan:2012xdj},
the precise determination of its properties is one of the mayor goals of
run II of the LHC.  The expected experimental precision in the upcoming
years has to be matched by theory predictions.  Since the dominant
production mechanism for Higgs boson production at the LHC is gluon
fusion, QCD corrections are the most sizable ones.  In fact, the
combined next-to-leading order~(NLO) and next-to-next-to-leading
order~(NNLO) corrections amount to the same size as the leading order
(LO) cross section.  The NNLO prediction still has to be assigned an
uncertainty of about $15\%$ which is to one part due to the parton
densities and to another part due to unknown higher orders in the
perturbation series of the partonic quantity.  Therefore, our interest
lies in the computation of the partonic cross section to
next-to-next-to-next-to-leading order~(N$^3$LO).

\subsection*{Effective field theory}

Gluon fusion is a process induced by quark loops and that is dominated
by the top quark.  Hence, it is reasonable to work within an effective
field theory~(EFT) where the top quark has been integrated out.  In
addition to the ordinary five-flavor Lagrangian of QCD, one has to
consider the interaction term
\begin{align}
  \mathcal{L}_{Y, \text{eff}} = -\frac{H}{v} C_1 \mathcal{O}_1, \quad
  \mathcal{O}_1 = \frac{1}{4} G_{\mu \nu}^a G^{a, \mu \nu},
\end{align}
where $H$ denotes the Higgs field, $v$ its vacuum expectation value and
$C_1$ the matching coefficient between full and effective theory.  The
effective coupling operator $\mathcal{O}_1$ is composed of the gluon
field strength tensor $G_{\mu \nu}$.

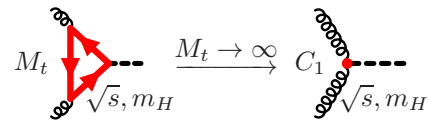
\begin{figure}[tb]
  \begin{fmffile}{LO_EFT}
    \begin{align*}
      \parbox{40pt}{
        \begin{fmfgraph*}(40,40)
          % style
          \fmfstyle
          \fmfcurved
          \fmfset{arrow_ang}{25}
          \fmfset{arrow_len}{3.0mm}
          % external vertices
          \fmfleft{i2,i1}
          \fmfright{o1}
          % external lines
          \fmf{curly}{v1,i1}
          \fmf{curly}{i2,v2}
          \fmf{dashes,w=2.0,la=$\sqrt{s},,m_H$,l.d=8.0}{v3,o1}
          % internal lines
          \fmf{plain_arrow,f=red,w=2.5,t=0.5}{v2,v3}
          \fmf{plain_arrow,f=red,w=2.5,t=0.5}{v3,v1}
          \fmffreeze
          \fmf{plain_arrow,f=red,w=2.5,la=$M_t$,l.d=8.0}{v1,v2}
        \end{fmfgraph*}}
      \quad \raisebox{5.0pt}{$\underrightarrow{M_t \to \infty}$} \quad
      \parbox{40pt}{
        \begin{fmfgraph*}(40,40)
          % style
          \fmfstyle
          \fmfcurved
          % external vertices
          \fmfleft{i2,i1}
          \fmfright{o1}
          % external lines
          \fmf{curly}{v1,i1}
          \fmf{curly}{i2,v1}
          \fmf{dashes,w=2.0,la=$\sqrt{s},,m_H$,l.d=8.0}{v1,o1}
          % vertices
          \fmfv{d.sh=circle,d.f=full,d.si=7.0thick,f=red,l=$C_1$,l.d=8.0}{v1}
        \end{fmfgraph*}}
    \end{align*}
    \vspace*{-1.0pc}
  \end{fmffile}
  \caption{%
    Contraction of the top quark loop corresponds to the transition from
    the full to the effective theory or to the limit $M_t \to \infty$.
    Thick straight lines represent the top quark, dashed lines the Higgs
    boson, dots its effective coupling and curly lines gluons.  Also, we
    indicated all relevant scales.}
  \label{fig:EFT}
\end{figure}

The transition from the full to the effective theory can be depicted
diagrammatically, see Fig.~\ref{fig:EFT}.

The advantage of working within the EFT setup is the smaller number of
diagrams and their reduced complexity.  There are less loops and less
scales that have to be considered simultaneously, as the dependence on
the top quark mass $M_t$ is encoded completely in $C_1$.  EFT diagrams
therefore depend only on a single dimensionless variable $x = m_H^2/s$
which is the squared ratio of the Higgs boson mass $m_H$ and the
partonic center-of-mass energy $\sqrt{s}$.

For Higgs boson production to N$^3$LO, the finite matching coefficient
$C_1$ is needed to four-loop order, see
Refs.~\cite{Chetyrkin:1997un,Schroder:2005hy,Chetyrkin:2005ia}.  The
renormalization of $C_1$ is described in Ref.~\cite{Spiridonov:1984br}
and can be expressed via the renormalization constant of $\alpha_s$
which is also itself needed to three-loop order, see
Refs.~\cite{Tarasov:1980au,Larin:1993tp}.

\subsection*{Optical theorem and Cutkosky rules}

The computation of higher order corrections to the cross section of a
process requires taking into account virtual (additional loops) and real
(additional final state particles) contributions.  Instead of
considering different phase spaces for the final states, one can employ
the optical theorem which relates the total cross section to the
imaginary part of the forward scattering amplitude.  Symbolically:
\begin{multline}
  \sigma\!\left( i \to f \right)
  \: \sim \:
  \sum_f \int \d{\Pi_f} \:
  \left| \mathcal{M}\!\left( i \to f \right) \right|^2 \\
  \: \sim \:
  \Im \mathcal{M}\!\left( i \to i \right),
\end{multline}
where $\sigma$ denotes the total cross section for an inital state $i$
to result in a collection of final states $f$ and $\mathcal{M}$ stands
for the matrix element for a transition between states (which are
identical for forward scattering).  The Cutkosky rules in turn relate
different contributions to the imaginary part of an amplitude to
specific sets of propagators which are set on-shell and thereby ``cut''.

\begin{figure}[tb]
  \begin{fmffile}{NNLO_opt-theo}
    \begin{multline*}
      \begin{aligned}
        & \int \d{\Pi_1} \left|
          \parbox{25pt}{
            \fmfframe(0,5)(0,5){
              \begin{fmfgraph*}(25,25)
                % style
                \fmfstyle
                \fmfcurved
                % external vertices
                \fmfleft{i2,i1}
                \fmfright{o1}
                % external lines
                \fmf{curly}{v1,i1}
                \fmf{curly}{i2,v1}
                \fmf{dashes,w=2.0}{v1,o1}
                % vertices
                \fmfv{d.sh=circle,d.f=full,d.si=5.0thick,f=red}{v1}
              \end{fmfgraph*}}}
          +
          \parbox{35pt}{
            \fmfframe(0,5)(0,5){
              \begin{fmfgraph*}(35,25)
                % style
                \fmfstyle
                \fmfcurved
                % external vertices
                \fmfleft{i2,i1}
                \fmfright{o1}
                % external lines
                \fmf{curly,t=2.0}{v1,i1}
                \fmf{curly,t=2.0}{i2,v2}
                \fmf{dashes,w=2.0,t=1.5}{v3,o1}
                % internal lines
                \fmf{curly}{v3,v1}
                \fmf{curly}{v2,v3}
                \fmffreeze
                \fmf{curly}{v1,v2}
                % vertices
                \fmfv{d.sh=circle,d.f=full,d.si=5.0thick,f=red}{v3}
              \end{fmfgraph*}}}
          + \ldots \; \right|^2 \\
        & \quad + \int \d{\Pi_2} \left|
          \parbox{25pt}{
            \fmfframe(0,5)(0,5){
              \begin{fmfgraph*}(25,25)
                % style
                \fmfstyle
                \fmfcurved
                % external vertices
                \fmfleft{i2,i1}
                \fmfright{o2,o1}
                % external lines
                \fmf{curly}{v1,i1}
                \fmf{curly}{i2,v2}
                \fmf{curly}{o1,v1}
                \fmf{dashes,w=2.0}{v2,o2}
                % internal lines
                \fmf{curly}{v1,v2}
                % vertices
                \fmfv{d.sh=circle,d.f=full,d.si=5.0thick,f=red}{v2}
              \end{fmfgraph*}}}
          +
          \parbox{30pt}{
            \fmfframe(0,5)(0,5){
              \begin{fmfgraph*}(30,25)
                % style
                \fmfstyle
                \fmfcurved
                % external vertices
                \fmfleft{i2,i1}
                \fmfright{o2,o1}
                % external lines
                \fmf{curly,t=2.0}{v1,i1}
                \fmf{curly,t=2.0}{i2,v2}
                \fmf{curly,t=2.0}{o1,v3}
                \fmf{dashes,w=2.0,t=2.0}{v4,o2}
                % internal lines
                \fmf{curly}{v3,v1}
                \fmf{curly}{v4,v3}
                \fmf{curly}{v2,v4}
                \fmf{curly}{v1,v2}
                % vertices
                \fmfv{d.sh=circle,d.f=full,d.si=5.0thick,f=red}{v4}
              \end{fmfgraph*}}}
          + \ldots \; \right|^2 \\
        & \quad + \int \d{\Pi_3} \left|
          \parbox{30pt}{
            \fmfframe(0,5)(0,5){
              \begin{fmfgraph*}(30,25)
                % style
                \fmfstyle
                \fmfcurved
                % external vertices
                \fmfleft{i2,i1}
                \fmfright{o3,o2,o1}
                % external lines
                \fmf{curly,t=2.0}{v1,i1}
                \fmf{curly,t=1.0}{i2,v2}
                \fmf{curly}{o1,v3}
                \fmf{curly,t=0.0}{v3,o2}
                \fmf{dashes,w=2.0}{v2,o3}
                % internal lines
                \fmf{curly}{v1,v2}
                \fmf{curly}{v3,v1}
                % vertices
                \fmfv{d.sh=circle,d.f=full,d.si=5.0thick,f=red}{v2}
              \end{fmfgraph*}}}
          +
          \parbox{35pt}{
            \fmfframe(0,5)(0,5){
              \begin{fmfgraph*}(35,25)
                % style
                \fmfstyle
                \fmfcurved
                % external vertices
                \fmfleft{i2,i1}
                \fmfright{o3,o2,o1}
                % external lines
                \fmf{curly,t=1.0}{v1,i1}
                \fmf{curly,t=2.0}{i2,v2}
                \fmf{curly}{o1,v1}
                \fmf{dashes,w=2.0,t=0.0}{v3,o2}
                \fmf{curly}{v3,o3}
                % internal lines
                \fmf{curly}{v1,v2}
                \fmf{curly}{v2,v3}
                % vertices
                \fmfv{d.sh=circle,d.f=full,d.si=5.0thick,f=red}{v3}
              \end{fmfgraph*}}}
          + \ldots \; \right|^2 + \ldots
      \end{aligned} \\
      \begin{aligned}
        = &
        \parbox{40pt}{
          \fmfframe(0,5)(0,5){
            \begin{fmfgraph*}(40,25)
              % style
              \fmfstyle
              \fmfcurved
              % external vertices
              \fmfleft{i2,i1}
              \fmfright{o2,o1}
              % external lines
              \fmf{curly}{v1,i1}
              \fmf{curly}{i2,v1}
              \fmf{curly}{o1,v2}
              \fmf{curly}{v2,o2}
              % internal lines
              \fmf{dashes,w=2.0}{v1,v2}
              % vertices
              \fmfv{d.sh=circle,d.f=full,d.si=5.0thick,f=red}{v1,v2}
              % cuts
              \fmffreeze
              \fmfdraw
              \fmfforce{(0.5w,-0.1h)}{c1}
              \fmfforce{(0.5w,1.1h)}{c2}
              \fmf{zigzag,rubout}{c1,c2}
            \end{fmfgraph*}}}
        +
        \parbox{40pt}{
          \fmfframe(0,5)(0,5){
            \begin{fmfgraph*}(40,25)
              % style
              \fmfstyle
              \fmfcurved
              % external vertices
              \fmfleft{i2,i1}
              \fmfright{o2,o1}
              % external lines
              \fmf{curly,t=1.5}{v1,i1}
              \fmf{curly,t=1.5}{i2,v2}
              \fmf{curly,t=1.5}{o1,v4}
              \fmf{curly,t=1.5}{v4,o2}
              % internal lines
              \fmf{curly}{v3,v1}
              \fmf{curly}{v2,v3}
              \fmf{dashes,w=2.0}{v3,v4}
              \fmffreeze
              \fmf{curly}{v1,v2}
              % vertices
              \fmfv{d.sh=circle,d.f=full,d.si=5.0thick,f=red}{v3,v4}
              % cuts
              \fmffreeze
              \fmfdraw
              \fmfforce{(0.63w,-0.1h)}{c1}
              \fmfforce{(0.63w,1.1h)}{c2}
              \fmf{zigzag,rubout}{c1,c2}
            \end{fmfgraph*}}}
        + \ldots +
        \parbox{30pt}{
          \fmfframe(0,5)(0,5){
            \begin{fmfgraph*}(30,25)
              % style
              \fmfstyle
              \fmfcurved
              % external vertices
              \fmfleft{i2,i1}
              \fmfright{o2,o1}
              % external lines
              \fmf{curly,t=2.5}{v1,i1}
              \fmf{curly,t=2.5}{i2,v2}
              \fmf{curly,t=2.5}{o1,v3}
              \fmf{curly,t=2.5}{v4,o2}
              % internal lines
              \fmf{curly}{v3,v1}
              \fmf{curly}{v4,v3}
              \fmf{dashes,w=2.0}{v2,v4}
              \fmf{curly}{v1,v2}
              % vertices
              \fmfv{d.sh=circle,d.f=full,d.si=5.0thick,f=red}{v2,v4}
              % cuts
              \fmffreeze
              \fmfdraw
              \fmfforce{(0.5w,-0.1h)}{c1}
              \fmfforce{(0.5w,1.1h)}{c2}
              \fmf{zigzag,rubout}{c1,c2}
            \end{fmfgraph*}}} + \ldots \\
        & \qquad +
        \parbox{50pt}{
          \fmfframe(0,5)(0,5){
            \begin{fmfgraph*}(50,25)
              % style
              \fmfstyle
              \fmfcurved
              % external vertices
              \fmfleft{i2,i1}
              \fmfright{o2,o1}
              % external lines
              \fmf{curly,t=1.5}{v1,i1}
              \fmf{curly,t=1.5}{i2,v2}
              \fmf{curly,t=1.5}{o1,v3}
              \fmf{curly,t=1.5}{v4,o2}
              % internal lines
              \fmf{curly,t=0.5}{v5,v1}
              \fmf{curly,t=0.5}{v3,v5}
              \fmf{curly,t=0.5}{v4,v3}
              \fmf{dashes,w=2.0,t=0.5}{v6,v4}
              \fmf{curly,t=0.5}{v2,v6}
              \fmf{curly,t=0.5}{v1,v2}
              \fmffreeze
              \fmf{curly,t=0.5}{v5,v6}
              % vertices
              \fmfv{d.sh=circle,d.f=full,d.si=5.0thick,f=red}{v4,v6}
              % cuts
              \fmffreeze
              \fmfdraw
              \fmfforce{(0.65w,-0.1h)}{c1}
              \fmfforce{(0.65w,1.1h)}{c2}
              \fmf{zigzag,rubout}{c1,c2}
            \end{fmfgraph*}}}
        + \ldots +
        \parbox{50pt}{
          \fmfframe(0,5)(0,5){
            \begin{fmfgraph*}(50,25)
              % style
              \fmfstyle
              \fmfcurved
              % external vertices
              \fmfleft{i2,i1}
              \fmfright{o2,o1}
              % external lines
              \fmf{curly,t=1.5}{v1,i1}
              \fmf{curly,t=1.5}{i2,v2}
              \fmf{curly,t=1.5}{o1,v3}
              \fmf{curly,t=1.5}{v4,o2}
              % internal lines
              \fmf{curly,t=0.5}{v5,v1}
              \fmf{curly,t=0.5}{v3,v5}
              \fmf{curly,t=0.5}{v4,v3}
              \fmf{dashes,w=2.0,t=0.5}{v6,v4}
              \fmf{curly,t=0.5}{v2,v6}
              \fmf{curly,t=0.5}{v1,v2}
              \fmffreeze
              \fmf{curly,t=0.5}{v5,v6}
              % vertices
              \fmfv{d.sh=circle,d.f=full,d.si=5.0thick,f=red}{v4,v6}
              % cuts
              \fmffreeze
              \fmfdraw
              \fmfforce{(0.65w,-0.1h)}{c1}
              \fmfforce{(0.65w,0.5h)}{c2}
              \fmfforce{(0.35w,0.5h)}{c3}
              \fmfforce{(0.35w,1.1h)}{c4}
              \fmf{zigzag,rubout}{c1,c2,c3,c4}
            \end{fmfgraph*}}}
        + \ldots
      \end{aligned}
    \end{multline*}
  \end{fmffile}
  \vspace*{-1.0pc}
  \caption{%
    Illustration of the optical theorem up NNLO, see the main text for
    details.  Apart from the notation in Fig.~\ref{fig:EFT}, wiggly
    lines represent cuts.}
  \label{fig:opt-theo}
\end{figure}

Figure~\ref{fig:opt-theo} exemplifies this technique up to NNLO.  The
left-hand side shows the phase space integrals over squared production
amplitudes for the Higgs boson with up to two additional partons.  The
right-hand side shows the corresponding interference diagrams where the
cut-line is equivalent to a phase space.  Note, the last two diagrams
are different cuts of the same amplitude.  This demonstrates the
validity just to consider all relevant cuts of the forward scattering
amplitude.

Working in forward scattering kinematics simplifies the calculation.
The optical theorem allows for common treatment of loop and phase space
integrals. Thus, the calculation of imaginary parts has to be performed
only for a relatively small set of ``master integrals'' (the integrals
remaining after application of integration-by-parts identities with
Laporta's algorithm).  This approach was first used in
Ref.~\cite{Anastasiou:2002yz} for the NNLO computation of Higgs
production.  On the other hand, within this method more diagrams with
more loops have to be computed and one has only access to a total
inclusive cross section, at least in its na\"ive application.

\subsection*{Existing results}

\begin{figure}[tb]
  \begin{fmffile}{N3LO_cuts}
    % -- VV2 --
    \parbox{60pt}{
      \centering
      \fmfframe(0,5)(0,25){
        \begin{fmfgraph*}(60,30)
          \fmfkeep{VV2}
          % style
          \fmfstyle
          \fmfcurved
          % external vertices
          \fmfleft{i2,i1}
          \fmfright{o2,o1}
          % external lines
          \fmf{curly,t=3.0}{v1,i1}
          \fmf{curly,t=3.0}{i2,v2}
          \fmf{curly,t=3.0}{o1,v3}
          \fmf{curly,t=3.0}{v4,o2}
          % internal lines
          \fmf{curly,t=2.0}{v6,v1}
          \fmf{curly,t=2.0}{v7,v6}
          \fmf{curly}{v2,v7}
          \fmf{dashes,w=2.0,t=3.0}{v7,v8}
          \fmf{curly,t=2.0}{v3,v8}
          \fmf{curly,t=2.0}{v8,v4}
          \fmffreeze
          \fmf{curly}{v5,v2}
          \fmf{curly}{v1,v5}
          \fmf{curly}{v4,v3}
          \fmffreeze
          \fmf{curly}{v5,v6}
          % vertices
          \fmfv{d.sh=circle,d.f=full,d.si=5.0thick,f=red}{v7,v8}
          % cuts
          \fmffreeze
          \fmfdraw
          \fmfforce{(0.57w,-0.1h)}{c1}
          \fmfforce{(0.57w,1.1h)}{c2}
          \fmf{zigzag,rubout}{c1,c2}
          \fmflabel{VV$^2$}{c1}
        \end{fmfgraph*}}}
    \hfill
    % -- V3 --
    \parbox{60pt}{
      \centering
      \fmfframe(0,5)(0,25){
        \begin{fmfgraph*}(60,40)
          \fmfkeep{V3}
          % style
          \fmfstyle
          \fmfcurved
          % external vertices
          \fmfleft{i2,i1}
          \fmfright{o2,o1}
          % external lines
          \fmf{curly,t=2.0}{v1,i1}
          \fmf{curly,t=2.0}{i2,v2}
          \fmf{curly,t=2.0}{o1,v3}
          \fmf{curly,t=2.0}{v3,o2}
          % internal lines
          \fmf{curly,t=1.3}{v4,v1}
          \fmf{curly}{v5,v4}
          \fmf{curly}{v6,v5}
          \fmf{curly,t=1.3}{v2,v6}
          \fmf{dashes,w=2.0,t=3.0}{v3,v5}
          \fmffreeze
          \fmf{curly}{v7,v2}
          \fmf{curly}{v1,v7}
          \fmffreeze
          \fmf{curly}{v4,v8}
          \fmf{curly}{v8,v6}
          \fmffreeze
          \fmf{curly}{v8,v7}
          % vertices
          \fmfv{d.sh=circle,d.f=full,d.si=5.0thick,f=red}{v3,v5}
          % cuts
          \fmffreeze
          \fmfdraw
          \fmfforce{(0.7w,-0.1h)}{c1}
          \fmfforce{(0.7w,1.1h)}{c2}
          \fmf{zigzag,rubout}{c1,c2}
          \fmflabel{\hspace*{-0.5pc}V$^3$}{c1}
        \end{fmfgraph*}}}
    \hfill
    % -- VRV --
    \parbox{60pt}{
      \centering
      \fmfframe(0,5)(0,25){
        \begin{fmfgraph*}(60,30)
          \fmfkeep{VRV}
          % style
          \fmfstyle
          % external vertices
          \fmfleft{i2,i1}
          \fmfright{o2,o1}
          % external lines
          \fmf{curly,t=2.0}{i1,v1}
          \fmf{curly,t=2.0}{v2,i2}
          \fmf{curly,t=2.0}{v3,o1}
          \fmf{curly,t=2.0}{o2,v4}
          % internal lines
          \fmf{curly}{v4,v3}
          \fmf{curly}{v1,v2}
          \fmffreeze
          \fmf{curly}{v5,v1}
          \fmf{curly}{v6,v5}
          \fmf{curly}{v3,v6}
          \fmf{curly}{v8,v4}
          \fmf{dashes,w=2.0}{v7,v8}
          \fmf{curly}{v2,v7}
          \fmffreeze
          \fmf{curly}{v5,v7}
          \fmf{curly}{v8,v6}
          % vertices
          \fmfv{d.sh=circle,d.f=full,d.si=5.0thick,f=red}{v7,v8}
          % cuts
          \fmffreeze
          \fmfdraw
          \fmfforce{(0.5w,-0.1h)}{c1}
          \fmfforce{(0.5w,1.1h)}{c2}
          \fmf{zigzag,rubout}{c1,c2}
          \fmflabel{VRV}{c1}
        \end{fmfgraph*}}}
    \\
    % -- V2R --
    \parbox{60pt}{
      \centering
      \fmfframe(0,5)(0,25){
        \begin{fmfgraph*}(60,40)
          \fmfkeep{V2R}
          % style
          \fmfstyle
          % external vertices
          \fmfleft{i2,i1}
          \fmfright{o2,o1}
          % external lines
          \fmf{curly,t=3.0}{v1,i1}
          \fmf{curly,t=3.0}{i2,v2}
          \fmf{curly,t=3.0}{o1,v3}
          \fmf{curly,t=3.0}{v4,o2}
          % internal lines
          \fmf{curly}{v5,v1}
          \fmf{curly,t=1.5}{v3,v5}
          \fmf{curly,t=0.5}{v4,v3}
          \fmf{dashes,w=2.0,t=1.5}{v4,v6}
          \fmf{curly}{v2,v6}
          \fmf{curly}{v7,v2}
          \fmf{curly}{v1,v7}
          \fmffreeze
          \fmf{curly}{v8,v5}
          \fmf{curly}{v6,v8}
          \fmf{curly,t=0.0}{v8,v7}
          % vertices
          \fmfv{d.sh=circle,d.f=full,d.si=5.0thick,f=red}{v4,v6}
          % cuts
          \fmffreeze
          \fmfdraw
          \fmfforce{(0.67w,-0.1h)}{c1}
          \fmfforce{(0.67w,1.1h)}{c2}
          \fmf{zigzag,rubout}{c1,c2}
          \fmflabel{\hspace*{-0.8pc}V$^2$R}{c1}
        \end{fmfgraph*}}}
    \hfill
    % -- VR2 --
    \parbox{60pt}{
      \centering
      \fmfframe(0,5)(0,25){
        \begin{fmfgraph*}(60,40)
          \fmfkeep{VR2}
          % style
          \fmfstyle
          % external vertices
          \fmfleft{i2,i1}
          \fmfright{o2,o1}
          % external lines
          \fmf{curly,t=3.0}{v1,i1}
          \fmf{curly,t=3.0}{i2,v2}
          \fmf{curly,t=3.0}{o1,v3}
          \fmf{curly,t=3.0}{v4,o2}
          % internal lines
          \fmf{curly,t=1.5}{v5,v1}
          \fmf{curly}{v3,v5}
          \fmf{curly}{v8,v3}
          \fmf{curly}{v4,v8}
          \fmf{dashes,w=2.0}{v4,v6}
          \fmf{curly,t=1.5}{v2,v6}
          \fmf{curly,t=0.5}{v1,v2}
          \fmffreeze
          \fmf{curly}{v5,v7}
          \fmf{curly}{v7,v6}
          \fmf{curly,t=0.0}{v8,v7}
          % vertices
          \fmfv{d.sh=circle,d.f=full,d.si=5.0thick,f=red}{v4,v6}
          % cuts
          \fmffreeze
          \fmfdraw
          \fmfforce{(0.62w,-0.1h)}{c1}
          \fmfforce{(0.62w,1.1h)}{c2}
          \fmf{zigzag,rubout}{c1,c2}
          \fmflabel{VR$^2$}{c1}
        \end{fmfgraph*}}}
    \hfill
    % -- R3 --
    \parbox{60pt}{
      \centering
      \fmfframe(0,5)(0,25){
        \begin{fmfgraph*}(40,50)
          \fmfkeep{R3}
          % style
          \fmfstyle
          % external vertices
          \fmfleft{i2,i1}
          \fmfright{o2,o1}
          % external lines
          \fmf{curly,t=2.0}{v1,i1}
          \fmf{curly,t=2.0}{i2,v2}
          \fmf{curly,t=2.0}{o1,v3}
          \fmf{curly,t=2.0}{v4,o2}
          % internal lines
          \fmf{curly}{v3,v1}
          \fmf{dashes,w=2.0}{v2,v4}
          \fmffreeze
          \fmf{curly}{v6,v3}
          \fmf{curly}{v8,v6}
          \fmf{curly}{v4,v8}
          \fmf{curly}{v7,v2}
          \fmf{curly}{v5,v7}
          \fmf{curly}{v1,v5}
          \fmffreeze
          \fmf{curly}{v6,v5}
          \fmf{curly}{v8,v7}
          % vertices
          \fmfv{d.sh=circle,d.f=full,d.si=5.0thick,f=red}{v2,v4}
          % cuts
          \fmffreeze
          \fmfdraw
          \fmfforce{(0.5w,-0.1h)}{c1}
          \fmfforce{(0.5w,1.1h)}{c2}
          \fmf{zigzag,rubout}{c1,c2}
          \fmflabel{R$^3$}{c1}
        \end{fmfgraph*}}}
    \\
    % --
  \end{fmffile}
  \vspace*{-1.0pc}
  \caption{%
    Different types of cut contributions at N$^3$LO.  The notation is as
    in Fig.~\ref{fig:opt-theo}, ``V'' stands for virtual and ``R'' for
    real in order to abbreviate contributions.}
  \label{fig:cuts}
\end{figure}
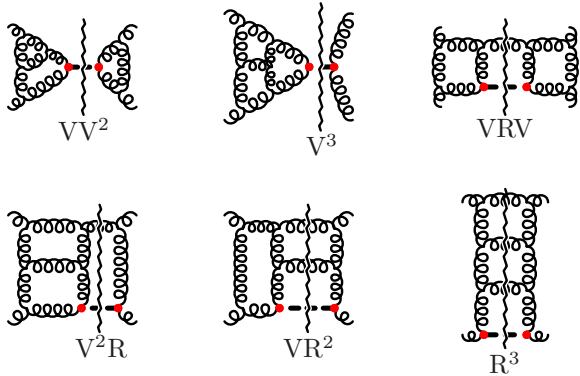

Let us give a listing of available results related to the Higgs boson
production cross section.  In Fig.~\ref{fig:cuts} we give sample
diagrams for the contributions from different cuts at N$^3$LO.

\begin{itemize}

\item The LO cross section (with exact dependence on $M_t$) was computed
  already in the seventies, see
  Refs.~\cite{Wilczek:1977zn,Ellis:1979jy,Georgi:1977gs,Rizzo:1979mf}.

\item NLO corrections (also exact in $M_t$) are available for almost
  twenty years, see Refs.~\cite{Dawson:1990zj,Spira:1995rr}.

\item NNLO corrections were first calculated within the EFT in
  Refs.~\cite{Harlander:2002wh,Anastasiou:2002yz,Ravindran:2003um}.
  Note that in Ref.~\cite{Harlander:2002wh} the soft expansion $x \to 1$
  to high orders was used where already the third order proofed to be
  valid within an error of \order{1\%}.

\item In
  Refs.~\cite{Harlander:2009bw,Pak:2009bx,Harlander:2009mq,Pak:2009dg,Harlander:2009my,Pak:2011hs}
  the expansion in $m_H^2/M_t^2$ was performed to assess the error
  within the EFT framework which turned out to be of \order{1\%}.

\item Combining three-loop splitting
  functions~\cite{Moch:2004pa,Vogt:2004mw} with partonic cross sections
  to NNLO expanded to higher orders in the dimensional
  regulator~\cite{Pak:2011hs,Anastasiou:2012kq}, convolution integrals
  and infrared counterterms could be presented in
  \cite{Hoschele:2012xc,Buehler:2013fha,Hoeschele:2013gga}.  In
  Ref.~\cite{Buehler:2013fha} the N$^3$LO scale variation was
  constructed and found to be of \order{2\% - 8\%}.

\item The three-loop gluon form factor was computed in
  Refs.~\cite{Baikov:2009bg,Gehrmann:2010ue} which gives the VV$^2$ and
  V$^3$ contributions.

\item For VRV and V$^2$R contributions (both exact in $x$), see
  Refs.~\cite{Anastasiou:2013mca,Kilgore:2013gba} and
  Refs.~\cite{Dulat:2014mda,Duhr:2014nda}, respectively.

\item For VR$^2$ and R$^3$ contributions (both as expansion in $x \to
  1$), see Refs.~\cite{Anastasiou:2014vaa,Li:2014afw,Anastasiou:2015ema}
  and Ref.~\cite{Anastasiou:2013srw}, respectively.  In
  Refs.~\cite{Anastasiou:2015ema} more than thirty terms in the soft
  expansion were reached for the total cross section which is sufficient
  for phenomenological applications.

\item In Refs.~\cite{Hoschele:2014qsa,Anzai:2015wma} the
  $qq^\prime$-channel was considered and exact results could be obtained
  for its VR$^2$ and R$^3$ contributions.

\end{itemize}

\section{Calculational techniques}

Generally, there is a steep raise in complexity when going to higher
loop orders in a specific process.  Here, there is only $1$ diagram at
LO, $50$ at NLO, $2\,946$ at NNLO and $174\,938$ at N$^3$LO.  The
corresponding integrals are classified in ``topologies'' or Feynman
integral families which are then subject to a reduction algorithm
resulting in a set of master integrals.  For the real corrections to
Higgs boson production $1$ topology is needed at NLO, $11$ at NNLO and
more than $100$ at N$^3$LO.  The number of master integrals is $1$ at
NLO, $20$ at NNLO and more than $100$ at N$^3$LO.

Our computational setup is highly automated.  First, we use {\tt
  QGRAF}~\cite{Nogueira:1991ex} to generate and a private
filter~\cite{diss_Hoff} to select all relevant Feynman diagrams.  We
perform two independent calculations, using either the program {\tt
  exp}~\cite{Harlander:1997zb,Seidensticker:1999bb} or {\tt reg} to map
diagrams to topologies.  The reduction to scalar integrals is then
performed with {\tt FORM}~\cite{Vermaseren:2000nd,Kuipers:2012rf}.  The
reduction to master integrals is done with {\tt rows} (an in-house
Laporta algorithm) and {\tt FIRE}~\cite{Smirnov:2014hma}.  The reduction
tables can reach a size of up to ten gigabytes.

The private code {\tt TopoID}~\cite{Grigo:2014oqa,diss_Hoff} is used to
provide input in an automatic fashion for the aforementioned steps.
Based on the appearing diagrams, we define a topology as set of
propagators allowing for permutations or contractions of propagators and
(linear) transformations of loop momenta.  {\tt TopoID} can also handle
topologies with linearly dependent propagators which are needed to map
all appearing diagrams.  For the partial fractioning relations and the
decomposition into topologies with linearly independent propagtors {\tt
  FORM} code is generated.  Since the momentum space representation of
topologies is ambiguous, we employ the Feynman representation in a
unique form to identify duplicate topologies and also to find a minimal
basis of master integrals in the very end.  This form of the Feynman
representation is also used to obtain all possible symmetries of
topologies which is an enormous aid in the Laporta reduction.  Moreover,
{\tt TopoID} is able to handle cuts of Feynman diagrams in the sense
that all possible ways are detected to set propagators on-shell giving
rise to an imaginary part.  For details we refer the interesed reader
to~\cite{diss_Hoff}.

The master integrals are calculated using the technique of canonical
differential equations~(DEQs), see e.g. Ref.~\cite{Henn:2013pwa}.  In
general, one obtains a coupled system of linear first-order DEQs for the
master integrals.  This system is generated by applying the derivative
in a kinematic invariant or mass on the integrand of a master integral
and using the reduction procedure subsequently.  We refrain from giving
details on the computation of the soft limit $x \to 1$ which is used as
boundary condition and refer instead to Ref.~\cite{Anzai:2015wma}.

In the last few years Henn~\cite{Henn:2013pwa}, advertised a specific
form of DEQs which can be reached via a basis transformation of the
master integrals and cast in the form
\begin{align}
  \frac{\d{}}{\d{x}} m_i\!\left( x, \ep \right) =
  \ep \, A_{ij}\!\left( x \right) \, m_j\!\left( x, \ep \right),
\end{align}
where $m_i$ is a vector of master integrals, $A_{i j}$ the fundamental
matrix of the DEQs, $x$ a scaleless variable and $d = 4 - 2 \, \ep$ the
number of spacetime dimensions.  The appeal of this form lies in the
factorization of the dependence on the dimensional regulator $\ep$ and
the kinematics.  In this form the system can be solved order by order in
$\ep$ and one can immediately read off the so-called ``alphabet'', the
set of integration kernels, of the iterated integrals that pose the
solutions.

Up to NNLO all integrals can be represented as Harmonic
Polylogarithms~(HPLs), denoted $\HPL{\vec w}$, which are defined as
follows:
\begin{align}
  \begin{aligned}
    & \HPL{\vec w} =
    \int_0^x \d{x'} \:
    f_{w_1}\!\left( x' \right) \HPL[x']{\vec w_{n - 1}}, \\
    & f_0\!\left( x \right) = \frac{1}{x}, \;
    f_1\!\left( x \right) = \frac{1}{1 - x}, \;
    f_{-1}\!\left( x \right) = \frac{1}{1 + x},
  \end{aligned}
  \label{eq:HPL}
\end{align}
where $\vec w = \left( w_1, \vec w_{n - 1} \right)$ is the vector of $n$
weights and the $f_i$ are the integration kernels.  On N$^3$LO, as we
will see in Section~\ref{sec:qp}, HPLs are not sufficient anymore and
the class of functions needs to be extended.

\section{The $qq^\prime$-channel}
\label{sec:qp}

\begin{figure}[tb]
  \begin{fmffile}{N3LOqp_tops}
    \begin{center}
      % -- N3LOqpBT1 --
      \parbox{60pt}{
        \fmfframe(0,5)(0,5){
          \begin{fmfgraph}(60,30)
            \fmfstyle
            % cuts
            \fmffreeze
            \fmfdraw
            \fmfforce{(0.5w,-0.1h)}{c1}
            \fmfforce{(0.5w,1.1h)}{c2}
            \fmf{zigzag,w=1.0,fore=(0.8,,0.8,,0.8)}{c1,c2}
            % external vertices
            \fmfleft{i2,i1}
            \fmfright{o4,o3}
            % external lines
            \fmf{plain_arrow,t=3.0}{i1,v1}
            \fmf{plain,t=3.0}{i2,v2}
            \fmf{plain_arrow,t=3.0}{v3,o3}
            \fmf{plain,t=3.0}{v4,o4}
            % internal lines
            \fmf{plain,t=2.0,l.d=4.0}{v7,v4}
            \fmf{plain,t=1.0,l.d=4.0}{v2,v7}
            \fmf{plain,t=1.0,l.d=4.0}{v3,v5}
            \fmf{plain,t=2.0,l.d=4.0}{v5,v1}
            \fmffreeze
            \fmf{dbl_plain,t=0.0,l.d=4.0}{v8,v6}
            \fmf{plain,t=1.0,l.d=4.0}{v7,v8}
            \fmf{plain,t=1.0,l.d=4.0}{v6,v4}
            \fmf{plain,t=2.0,l.d=4.0}{v6,v1}
            \fmf{plain,t=2.0,l.d=4.0}{v8,v3}
            \fmf{plain,t=1.0,l.d=4.0}{v5,v2}
          \end{fmfgraph}}}
      % -- N3LOqpBT2 --
      \parbox{60pt}{
        \fmfframe(0,5)(0,5){
          \begin{fmfgraph*}(60,30)
            \fmfstyle
            % cuts
            \fmfforce{(0.73w,-0.1h)}{c1}
            \fmfforce{(0.73w,1.1h)}{c2}
            \fmf{zigzag,w=1.0,fore=(0.7,,0.7,,0.7)}{c1,c2}
            \fmfforce{(0.4w,-0.1h)}{c3}
            \fmfforce{(0.57w,1.1h)}{c4}
            \fmf{zigzag,w=1.0,fore=(0.7,,0.7,,0.7)}{c3,c4}
            \fmfforce{(0.57w,-0.1h)}{c5}
            \fmfforce{(0.4w,1.1h)}{c6}
            \fmf{zigzag,w=1.0,fore=(0.7,,0.7,,0.7)}{c5,c6}
            % external vertices
            \fmfleft{i2,i1}
            \fmfright{o4,o3}
            % external lines
            \fmf{plain_arrow,t=3.0}{i1,v1}
            \fmf{plain,t=3.0}{i2,v3}
            \fmf{plain_arrow,t=3.0}{v4,o3}
            \fmf{plain,t=3.0}{v2,o4}
            % internal lines
            \fmf{plain,t=1.0,l.d=4.0}{v4,v6}
            \fmf{plain,t=1.0,l.d=4.0}{v6,v1}
            \fmf{plain,t=1.0,l.d=4.0}{v3,v5}
            \fmf{plain,t=1.0,l.d=4.0}{v5,v2}
            \fmffreeze
            \fmf{dbl_plain,t=0.0,l.d=4.0}{v8,v7}
            \fmf{plain,t=1.0,l.d=0.0}{v6,v8}
            \fmf{plain,t=1.0,l.d=4.0}{v7,v4}
            \fmf{plain,t=0.5,l.d=4.0}{v2,v7}
            \fmf{plain,t=0.5,l.d=4.0}{v3,v8}
            \fmf{plain,t=0.0,l.d=0.0}{v1,v5}
      \end{fmfgraph*}}}
      % -- N3LOqpBT3 --
      \parbox{60pt}{
        \fmfframe(0,5)(0,5){
          \begin{fmfgraph*}(60,30)
            \fmfstyle
            % cuts
            \fmffreeze
            \fmfdraw
            \fmfforce{(0.75w,-0.1h)}{c1}
            \fmfforce{(0.2w,1.1h)}{c2}
            \fmf{zigzag,w=1.0,fore=(0.7,,0.7,,0.7),r=0.15}{c1,c2}
            \fmfforce{(0.2w,-0.1h)}{c3}
            \fmfforce{(0.75w,1.1h)}{c4}
            \fmf{zigzag,w=1.0,fore=(0.7,,0.7,,0.7),r=0.15}{c3,c4}
            % external vertices
            \fmfleft{i2,i1}
            \fmfright{o4,o3}
            % external lines
            \fmf{plain_arrow,t=3.0}{i1,v1}
            \fmf{plain,t=3.0}{i2,v2}
            \fmf{plain_arrow,t=3.0}{v3,o3}
            \fmf{plain,t=3.0}{v4,o4}
            % internal lines
            \fmf{plain,t=1.0,l.d=4.0}{v3,v5}
            \fmf{plain,t=1.0,l.d=4.0}{v6,v4}
            \fmf{plain,t=1.0,l.d=4.0}{v5,v1}
            \fmf{plain,t=1.0,l.d=4.0}{v2,v6}
            \fmffreeze
            \fmf{dbl_plain,t=0.5,l.d=4.0}{v8,v7}
            \fmf{plain,t=1.0,l.d=4.0}{v5,v8}
            \fmf{plain,t=1.0,l.d=4.0}{v7,v6}
            \fmf{plain,t=0.0,l.d=1.5}{v7,v3}
            \fmf{plain,t=0.0,l.d=1.5}{v4,v8}
            \fmf{plain,t=0.0,l.d=4.0}{v1,v2}
      \end{fmfgraph*}}}
      % -- N3LOqpBT4 --
      \parbox{60pt}{
        \fmfframe(0,5)(0,5){
          \begin{fmfgraph*}(60,30)
            \fmfstyle
            % cuts
            \fmfforce{(0.48w,-0.1h)}{c1}
            \fmfforce{(0.7w,1.1h)}{c2}
            \fmf{zigzag,w=1.0,fore=(0.7,,0.7,,0.7)}{c1,c2}
            % external vertices
            \fmfleft{i2,i1}
            \fmfright{o4,o3}
            % external lines
            \fmf{plain_arrow,t=3.0}{i1,v1}
            \fmf{plain,t=3.0}{i2,v4}
            \fmf{plain_arrow,t=3.0}{v3,o3}
            \fmf{plain,t=3.0}{v2,o4}
            % internal lines
            \fmf{plain,t=1.0,l.d=4.0}{v4,v6}
            \fmf{plain,t=1.0,l.d=4.0}{v3,v5}
            \fmf{plain,t=2.0,l.d=4.0}{v6,v2}
            \fmf{plain,t=2.0,l.d=4.0}{v5,v1}
            \fmffreeze
            \fmf{dbl_plain,t=0.0,l.d=0.0}{v8,v7}
            \fmf{plain,t=1.0,l.d=0.0}{v6,v7}
            \fmf{plain,t=1.0,l.d=0.0}{v8,v5}
            \fmf{plain,t=1.0,l.d=4.0}{v8,v4}
            \fmf{plain,t=1.0,l.d=4.0}{v3,v7}
            \fmf{plain,t=0.0,l.d=0.0}{v2,v1}
      \end{fmfgraph*}}}
      % -- N3LOqpBT5 --
      \parbox{60pt}{
        \fmfframe(0,5)(0,5){
          \begin{fmfgraph*}(60,30)
            \fmfstyle
            % cuts
            \fmffreeze
            \fmfdraw
            \fmfforce{(0.73w,-0.1h)}{c1}
            \fmfforce{(0.73w,1.1h)}{c2}
            \fmf{zigzag,w=1.0,fore=(0.7,,0.7,,0.7)}{c1,c2}
            \fmfforce{(0.4w,-0.1h)}{c3}
            \fmfforce{(0.4w,0.1h)}{c4}
            \fmfforce{(0.6w,0.2h)}{c5}
            \fmfforce{(0.6w,1.1h)}{c6}
            \fmf{zigzag,w=1.0,fore=(0.7,,0.7,,0.7)}{c3,c4,c5,c6}
            \fmfforce{(0.55w,-0.1h)}{c7}
            \fmfforce{(0.55w,0.8h)}{c8}
            \fmfforce{(0.4w,0.9h)}{c9}
            \fmfforce{(0.4w,1.1h)}{c10}
            \fmf{zigzag,w=1.0,fore=(0.7,,0.7,,0.7)}{c7,c8,c9,c10}
            % external vertices
            \fmfleft{i2,i1}
            \fmfright{o4,o3}
            % external lines
            \fmf{plain_arrow,t=3.0}{i1,v1}
            \fmf{plain,t=3.0}{i2,v2}
            \fmf{plain_arrow,t=3.0}{v3,o3}
            \fmf{plain,t=3.0}{v4,o4}
            % internal lines
            \fmf{plain,t=1.0,l.d=4.0}{v3,v5}
            \fmf{plain,t=1.0,l.d=4.0}{v5,v1}
            \fmf{plain,t=1.0,l.d=4.0}{v2,v6}
            \fmf{plain,t=1.0,l.d=4.0}{v6,v4}
            \fmffreeze
            \fmf{dbl_plain,t=0.0,l.d=4.0}{v7,v8}
            \fmf{plain,t=0.1,l.d=4.0}{v5,v8}
            \fmf{plain,t=0.2,l.d=4.0}{v7,v3}
            \fmf{plain,t=0.2,l.d=4.0}{v4,v7}
            \fmf{plain,t=0.1,l.d=4.0}{v6,v8}
            \fmf{plain,t=0.2,l.d=4.0}{v1,v2}
      \end{fmfgraph*}}}
      % -- N3LOqpBT6 --
      \parbox{60pt}{
        \fmfframe(0,5)(0,5){
          \begin{fmfgraph*}(60,30)
            \fmfstyle
            % cuts
            \fmfforce{(0.65w,-0.1h)}{c1}
            \fmfforce{(0.35w,1.1h)}{c2}
            \fmf{zigzag,w=1.0,fore=(0.7,,0.7,,0.7),r=0.35}{c1,c2}
            % external vertices
            \fmfleft{i2,i1}
            \fmfright{o4,o3}
            % external lines
            \fmf{plain_arrow,t=1.0}{i1,v1}
            \fmf{plain,t=2.0}{i2,v3}
            \fmf{plain_arrow,t=2.0}{v4,o3}
            \fmf{plain,t=2.0}{v2,o4}
            % internal lines
            \fmf{plain,t=1.5,l.d=4.0}{v5,v2}
            \fmf{plain,t=2.0,l.d=4.0}{v4,v6}
            \fmf{plain,t=1.0,l.d=4.0}{v6,v1}
            \fmf{plain,t=2.5,l.d=4.0}{v3,v5}
            \fmffreeze
            \fmf{dbl_plain,t=1.0,l.d=4.0,r=0.5}{v7,v4}
            \fmf{plain,t=1.0,l.d=0.0,l=0.5}{v7,v4}
            \fmf{plain,t=1.5,l.d=4.0}{v3,v7}
            \fmf{plain,t=0.0,l.d=0.0}{v2,v6}
            \fmf{plain,t=0.0,l.d=4.0}{v1,v5}
      \end{fmfgraph*}}}
      % -- N3LOqpBT7 --
      \parbox{60pt}{
        \fmfframe(0,5)(0,5){
          \begin{fmfgraph*}(60,30)
            \fmfstyle
            % cuts
            \fmfforce{(0.43w,-0.1h)}{c1}
            \fmfforce{(0.43w,1.1h)}{c2}
            \fmf{zigzag,w=1.0,fore=(0.7,,0.7,,0.7)}{c1,c2}
            % external vertices
            \fmfleft{i2,i1}
            \fmfright{o4,o3}
            % external lines
            \fmf{plain_arrow,t=3.0}{i1,v3}
            \fmf{plain,t=3.0}{i2,v1}
            \fmf{plain_arrow,t=3.0}{v4,o3}
            \fmf{plain,t=3.0}{v2,o4}
            % internal lines
            \fmf{phantom,t=2,0}{v3,v0}
            \fmf{phantom,t=2.0}{v0,v4}
            \fmf{plain,t=1.0,l.d=4.0}{v1,v2}
            \fmffreeze
            \fmf{plain,t=1.0,l.d=4.0}{v2,v8}
            \fmf{plain,t=1.0,l.d=4.0}{v3,v5}
            \fmf{plain,t=1.0,l.d=4.0}{v8,v4}
            \fmf{plain,t=1.0,l.d=4.0}{v5,v1}
            \fmffreeze
            \fmf{dbl_plain,t=1.0,l.d=4.0}{v5,v7}
            \fmf{plain,t=1.0,l.d=0.0}{v7,v8}
            \fmf{plain,t=0.0,l.d=0.0,r=0.3}{v2,v3}
            \fmf{plain,t=0.0,l.d=4.0}{v4,v7}
      \end{fmfgraph*}}}
      % -- N3LOqpBT8 --
      \parbox{60pt}{
        \fmfframe(0,5)(0,5){
          \begin{fmfgraph*}(60,30)
            \fmfstyle
            % cuts
            \fmfforce{(0.52w,-0.1h)}{c1}
            \fmfforce{(0.37w,1.1h)}{c2}
            \fmf{zigzag,w=1.0,fore=(0.7,,0.7,,0.7)}{c1,c2}
            % external vertices
            \fmfleft{i2,i1}
            \fmfright{o4,o3}
            % external lines
            \fmf{plain_arrow,t=2.0}{i1,v1}
            \fmf{plain,t=2.0}{i2,v2}
            \fmf{plain_arrow,t=2.0}{v4,o3}
            \fmf{plain,t=2.0}{v3,o4}
            % internal lines
            \fmf{plain,t=1.0,l.d=4.0}{v6,v3}
            \fmf{plain,t=2.0,l.d=4.0}{v2,v6}
            \fmf{plain,t=2.0,l.d=0.0}{v5,v1}
            \fmf{plain,t=1.0,l.d=4.0}{v4,v5}
            \fmffreeze
            \fmf{dbl_plain,t=1.0,l.d=4.0,r=0.5}{v7,v6}
            \fmf{plain,t=1.0,l.d=0.0,l=0.5}{v7,v6}
            \fmf{plain,t=0.0,l.d=0.0}{v5,v3}
            \fmf{plain,t=1.5,l.d=4.0}{v4,v7}
            \fmf{plain,t=0.0,l.d=4.0}{v1,v2}
      \end{fmfgraph*}}}
      % -- N3LOqpBT9 --
      \parbox{60pt}{
        \fmfframe(0,5)(0,5){
          \begin{fmfgraph*}(60,30)
            \fmfstyle
            % cuts
            \fmfforce{(0.27w,-0.1h)}{c1}
            \fmfforce{(0.27w,1.1h)}{c2}
            \fmf{zigzag,w=1.0,fore=(0.7,,0.7,,0.7)}{c1,c2}
            \fmfforce{(0.72w,-0.1h)}{c3}
            \fmfforce{(0.72w,1.1h)}{c4}
            \fmf{zigzag,w=1.0,fore=(0.7,,0.7,,0.7)}{c3,c4}
            \fmfforce{(0.42w,-0.1h)}{c5}
            \fmfforce{(0.42w,0.1h)}{c6}
            \fmfforce{(0.6w,0.35h)}{c7}
            \fmfforce{(0.6w,1.1h)}{c8}
            \fmf{zigzag,w=1.0,fore=(0.7,,0.7,,0.7)}{c5,c6,c7,c8}
            \fmfforce{(0.6w,-0.1h)}{c9}
            \fmfforce{(0.6w,0.15h)}{c10}
            \fmfforce{(0.42w,0.4h)}{c11}
            \fmfforce{(0.42w,1.1h)}{c12}
            \fmf{zigzag,w=1.0,fore=(0.7,,0.7,,0.7)}{c9,c10,c11,c12}
            % external vertices
            \fmfleft{i2,i1}
            \fmfright{o4,o3}
            % external lines
            \fmf{plain_arrow,t=3.0}{i1,v1}
            \fmf{plain,t=3.0}{i2,v2}
            \fmf{plain_arrow,t=3.0}{v3,o3}
            \fmf{plain,t=3.0}{v4,o4}
            % internal lines
            \fmf{plain,t=1.0,l.d=4.0}{v6,v1}
            \fmf{plain,t=1.0,l.d=4.0}{v3,v6}
            \fmf{plain,t=1.0,l.d=4.0}{v7,v4}
            \fmf{plain,t=1.0,l.d=4.0}{v2,v7}
            \fmffreeze
            \fmf{dbl_plain,t=0.0,l.d=0.0}{v5,v8}
            \fmf{plain,t=1.0,l.d=4.0}{v8,v3}
            \fmf{plain,t=1.0,l.d=4.0}{v1,v5}
            \fmf{plain,t=1.0,l.d=4.0}{v4,v8}
            \fmf{plain,t=1.0,l.d=4.0}{v5,v2}
            \fmf{plain,t=0.0,l.d=0.0}{v6,v7}
      \end{fmfgraph*}}}
      % -- N3LOqpBT10 --
      \parbox{60pt}{
        \fmfframe(0,5)(0,5){
          \begin{fmfgraph*}(60,30)
            \fmfstyle
            % cuts
            \fmfforce{(0.25w,-0.1h)}{c1}
            \fmfforce{(0.75w,1.1h)}{c2}
            \fmf{zigzag,w=1.0,fore=(0.7,,0.7,,0.7),r=0.2}{c1,c2}
            \fmfforce{(0.75w,-0.1h)}{c3}
            \fmfforce{(0.25w,1.1h)}{c4}
            \fmf{zigzag,w=1.0,fore=(0.7,,0.7,,0.7),l=0.15}{c3,c4}
            % external vertices
            \fmfleft{i2,i1}
            \fmfright{o4,o3}
            % external lines
            \fmf{plain_arrow,t=3.0}{i1,v1}
            \fmf{plain,t=3.0}{i2,v2}
            \fmf{plain_arrow,t=3.0}{v3,o3}
            \fmf{plain,t=3.0}{v4,o4}
            % internal lines
            \fmf{plain,t=1.0,l.d=4.0}{v3,v5}
            \fmf{plain,t=1.0,l.d=4.0}{v6,v4}
            \fmf{plain,t=1.0,l.d=4.0}{v2,v6}
            \fmf{plain,t=1.0,l.d=4.0}{v5,v1}
            \fmffreeze
            \fmf{dbl_plain,t=0.05,l.d=4.0,r=0.5}{v7,v6}
            \fmf{plain,t=0.05,l.d=4.0,l=0.5}{v7,v6}
            \fmf{plain,t=0.4,l.d=4.0}{v4,v3}
            \fmf{plain,t=0.1,l.d=4.0}{v5,v7}
            \fmf{plain,t=0.4,l.d=4.0}{v1,v2}
      \end{fmfgraph*}}}
      % -- N3LOqpBT11 --
      \parbox{60pt}{
        \centering
        \fmfframe(0,5)(0,5){
          \begin{fmfgraph*}(60,30)
            \fmfstyle
            % cuts
            \fmfforce{(0.67w,-0.1h)}{c1}
            \fmfforce{(0.67w,1.1h)}{c2}
            \fmf{zigzag,w=1.0,fore=(0.7,,0.7,,0.7)}{c1,c2}
            % external vertices
            \fmfleft{i2,i1}
            \fmfright{o4,o3}
            % external lines
            \fmf{plain_arrow,t=3.0}{i1,v1}
            \fmf{plain,t=3.0}{i2,v2}
            \fmf{plain_arrow,t=3.0}{v3,o3}
            \fmf{plain,t=6.0}{v4,o4}
            % internal lines
            \fmf{plain,t=1.5,l.d=4.0}{v5,v8}
            \fmf{plain,t=6.0,l.d=4.0}{v8,v4}
            \fmf{plain,t=3.0,l.d=4.0}{v2,v5}
            \fmf{plain,t=1.0,l.d=4.0}{v3,v1}
            \fmffreeze
            \fmf{dbl_plain,t=1.0,l.d=4.0}{v7,v3}
            \fmf{plain,t=1.0,l.d=4.0}{v7,v5}
            \fmf{plain,t=0.0,l.d=4.0}{v8,v3}
            \fmf{plain,t=0.0,l.d=0.0}{v4,v7}
            \fmf{plain,t=0.0,l.d=4.0}{v1,v2}
      \end{fmfgraph*}}}
      % -- N3LOqpBT12 --
      \parbox{60pt}{
        \fmfframe(0,5)(0,5){
          \begin{fmfgraph*}(60,30)
            \fmfstyle
            % cuts
            \fmfforce{(0.67w,-0.1h)}{c1}
            \fmfforce{(0.67w,1.1h)}{c2}
            \fmf{zigzag,w=1.0,fore=(0.7,,0.7,,0.7)}{c1,c2}
            % external vertices
            \fmfleft{i2,i1}
            \fmfright{o4,o3}
            % external lines
            \fmf{plain_arrow,t=3.0}{i1,v1}
            \fmf{plain,t=3.0}{i2,v2}
            \fmf{plain_arrow,t=3.0}{v3,o3}
            \fmf{plain,t=3.0}{v4,o4}
            % internal lines
            \fmf{plain,t=1.5,l.d=4.0}{v5,v4}
            \fmf{plain,t=3.0,l.d=4.0}{v2,v5}
            \fmf{plain,t=1.0,l.d=4.0}{v3,v1}
            \fmffreeze
            \fmf{dbl_plain,t=1.0,l.d=4.0}{v8,v3}
            \fmf{plain,t=1.0,l.d=4.0}{v7,v3}
            \fmf{plain,t=0.0,l.d=0.0}{v8,v7}
            \fmf{plain,t=1.5,l.d=4.0}{v8,v5}
            \fmf{plain,t=1.5,l.d=4.0}{v4,v7}
            \fmf{plain,t=0.0,l.d=4.0}{v1,v2}
      \end{fmfgraph*}}}
      % -- N3LOqpBT13 --
      \parbox{60pt}{
        \fmfframe(0,5)(0,5){
          \begin{fmfgraph*}(60,30)
            \fmfstyle
            % cuts
            \fmfforce{(0.67w,-0.1h)}{c1}
            \fmfforce{(0.67w,1.1h)}{c2}
            \fmf{zigzag,w=1.0,fore=(0.7,,0.7,,0.7)}{c1,c2}
            \fmfforce{(0.5w,-0.1h)}{c3}
            \fmfforce{(0.62w,0.5h)}{c4}
            \fmfforce{(0.6w,1.1h)}{c5}
            \fmf{zigzag,w=1.0,fore=(0.7,,0.7,,0.7),r=0.1}{c3,c4,c5}
            % external vertices
            \fmfleft{i2,i1}
            \fmfright{o4,o3}
            % external lines
            \fmf{plain_arrow,t=3.0}{i1,v1}
            \fmf{plain,t=3.0}{i2,v3}
            \fmf{plain_arrow,t=3.0}{v2,o3}
            \fmf{plain,t=3.0}{v4,o4}
            % internal lines
            \fmf{plain,t=2.0,l.d=0.0}{v6,v7}
            \fmf{plain,t=4.0,l.d=0.0}{v7,v4}
            \fmf{plain,t=4.0,l.d=4.0}{v3,v6}
            \fmf{plain,t=1.0,l.d=4.0}{v2,v1}
            \fmffreeze
            \fmf{dbl_plain,t=1.0,l.d=0.0}{v8,v2}
            \fmf{plain,t=0.0,l.d=4.0}{v8,v7}
            \fmf{plain,t=1.0,l.d=4.0}{v8,v6}
            \fmf{plain,t=0.0,l.d=4.0}{v4,v2}
            \fmf{plain,t=0.0,l.d=4.0}{v1,v3}
      \end{fmfgraph*}}}
      % -- N3LOqpBT14 --
      \parbox{60pt}{
        \fmfframe(0,5)(0,5){
          \begin{fmfgraph*}(60,30)
            \fmfstyle
            % cuts
            \fmfforce{(0.65w,-0.1h)}{c1}
            \fmfforce{(0.65w,1.1h)}{c2}
            \fmf{zigzag,w=1.0,fore=(0.7,,0.7,,0.7)}{c1,c2}
            % external vertices
            \fmfleft{i2,i1}
            \fmfright{o4,o3}
            % external lines
            \fmf{plain_arrow,t=3.0}{i1,v1}
            \fmf{plain,t=3.0}{i2,v3}
            \fmf{plain_arrow,t=3.0}{v2,o3}
            \fmf{plain,t=3.0}{v4,o4}
            % internal lines
            \fmf{plain,t=1.5,l.d=4.0}{v7,v4}
            \fmf{plain,t=1.0,l.d=4.0}{v2,v1}
            \fmf{plain,t=3.0,l.d=4.0}{v3,v7}
            \fmffreeze
            \fmf{dbl_plain,t=0.0,l.d=0.0}{v7,v2}
            \fmf{plain,t=1.0,l.d=4.0}{v8,v2}
            \fmf{plain,t=1.0,l.d=4.0}{v4,v8}
            \fmf{plain,t=0.0,l.d=4.0}{v8,v5}
            \fmf{plain,t=1.0,l.d=4.0}{v1,v5}
            \fmf{plain,t=1.0,l.d=4.0}{v5,v3}
      \end{fmfgraph*}}}
      % -- N3LOqpBT15 --
      \parbox{60pt}{
        \fmfframe(0,5)(0,5){
          \begin{fmfgraph*}(60,30)
            \fmfstyle
            % cuts
            \fmfforce{(0.33w,-0.1h)}{c1}
            \fmfforce{(0.33w,1.1h)}{c2}
            \fmf{zigzag,w=1.0,fore=(0.7,,0.7,,0.7)}{c1,c2}
            \fmfforce{(0.48w,-0.1h)}{c3}
            \fmfforce{(0.6w,1.1h)}{c4}
            \fmf{zigzag,w=1.0,fore=(0.7,,0.7,,0.7),r=0.1}{c3,c4}
            % external vertices
            \fmfleft{i2,i1}
            \fmfright{o4,o3}
            % external lines
            \fmf{plain_arrow,t=3.0}{i1,v1}
            \fmf{plain,t=3.0}{i2,v3}
            \fmf{plain_arrow,t=3.0}{v2,o3}
            \fmf{plain,t=3.0}{v4,o4}
            % internal lines
            \fmf{plain,t=3.0,l.d=0.0}{v7,v8}
            \fmf{plain,t=3.0,l.d=0.0}{v3,v7}
            \fmf{plain,t=3.0,l.d=4.0}{v8,v4}
            \fmf{plain,t=1.0,l.d=4.0}{v2,v1}
            \fmffreeze
            \fmf{dbl_plain,t=0.0,l.d=4.0}{v8,v5}
            \fmf{plain,t=1.0,l.d=4.0}{v5,v3}
            \fmf{plain,t=0.0,l.d=4.0}{v2,v7}
            \fmf{plain,t=2.0,l.d=4.0}{v1,v5}
            \fmf{plain,t=0.0,l.d=4.0}{v4,v2}
      \end{fmfgraph*}}}
      % -- N3LOqpBT16 --
      \parbox{60pt}{
        \fmfframe(0,5)(0,5){
          \begin{fmfgraph*}(60,30)
            \fmfstyle
            % cuts
            \fmfforce{(0.67w,-0.1h)}{c1}
            \fmfforce{(0.67w,1.1h)}{c2}
            \fmf{zigzag,w=1.0,fore=(0.7,,0.7,,0.7)}{c1,c2}
            % external vertices
            \fmfleft{i2,i1}
            \fmfright{o4,o3}
            % external lines
            \fmf{plain_arrow,t=3.0}{i1,v1}
            \fmf{plain,t=3.0}{i2,v3}
            \fmf{plain_arrow,t=3.0}{v2,o3}
            \fmf{plain,t=3.0}{v4,o4}
            % internal lines
            \fmf{plain,t=2.0,l.d=4.0}{v3,v5}
            \fmf{plain,t=1.0,l.d=4.0}{v2,v1}
            \fmf{plain,t=2.0,l.d=4.0}{v5,v4}
            \fmffreeze
            \fmf{dbl_plain,t=1.0,l.d=4.0}{v2,v7}
            \fmf{plain,t=1.0,l.d=4.0}{v2,v8}
            \fmf{plain,t=0.0,l.d=4.0}{v7,v8}
            \fmf{plain,t=0.0,l.d=0.0}{v1,v5}
            \fmf{plain,t=1.0,l.d=4.0}{v7,v3}
            \fmf{plain,t=1.0,l.d=4.0}{v8,v4}
      \end{fmfgraph*}}}
      % -- N3LOqpBT17 --
      \parbox{60pt}{
        \fmfframe(0,5)(0,5){
          \begin{fmfgraph*}(60,30)
            \fmfstyle
            % cuts
            \fmfforce{(0.4w,-0.1h)}{c1}
            \fmfforce{(0.65w,1.1h)}{c2}
            \fmf{zigzag,w=1.0,fore=(0.7,,0.7,,0.7),r=0.2}{c1,c2}
            % external vertices
            \fmfleft{i2,i1}
            \fmfright{o4,o3}
            % external lines
            \fmf{plain_arrow,t=3.0}{i1,v1}
            \fmf{plain,t=3.0}{i2,v3}
            \fmf{plain_arrow,t=3.0}{v2,o3}
            \fmf{plain,t=3.0}{v4,o4}
            % internal lines
            \fmf{dbl_plain,t=0.0,l.d=0.0,l=0.5}{v6,v7}
            \fmf{plain,t=2.0,l.d=4.0}{v6,v7}
            \fmf{plain,t=4.0,l.d=4.0}{v7,v4}
            \fmf{plain,t=4.0,l.d=4.0}{v3,v6}
            \fmf{plain,t=1.0,l.d=4.0}{v2,v1}
            \fmffreeze
            \fmf{plain,t=0.0,l.d=4.0}{v4,v2}
            \fmf{plain,t=0.0,l.d=4.0}{v6,v2}
            \fmf{plain,t=0.0,l.d=4.0}{v3,v1}
      \end{fmfgraph*}}}
      % --
    \end{center}
  \end{fmffile}
  \vspace*{-1.0pc}
  \caption{%
    The $17$ topologies used for the calculation of the
    $qq^\prime$-channel.  Plain lines are massless and double lines
    massive.  Arrows indicate the planar flow of external momenta and
    gray wiggly lines the three- and four-particle cuts.}
  \label{fig:qp}
\end{figure}
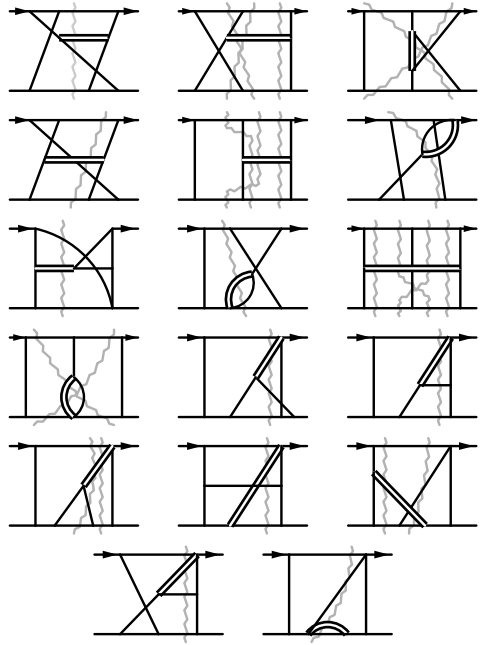

\begin{figure}[tb]
  \begin{center}
    \includegraphics[scale=0.5]{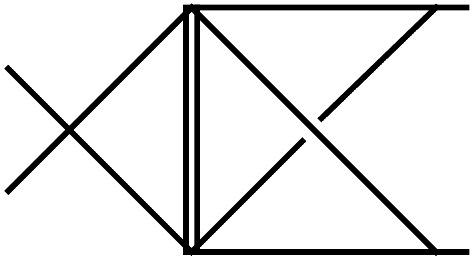}
  \end{center}
  \vspace*{-1.0pc}
  \caption{%
    Common graph of all integrals that give rise to functions beyond
    HPLs.}
  \label{fig:BT3}
\end{figure}

There are $220$ diagrams to be calculated in the $qq^\prime$-channel
which we chose to classify into $17$ topologies, see Fig.~\ref{fig:qp}.
Some of the integrals of only one of these topologies (the third in the
first row if Fig.~\ref{fig:qp}) could not be solved in terms of HPLs
only.  In that case the extended alphabet compared to Eq.~\eqref{eq:HPL}
yields also the ``letters''
\begin{align}
  f_{-4} = \frac{1}{1 + 4 \, x}, \;
  f_{s4} = \frac{1}{x} \left( \frac{1}{\sqrt{1 + 4 \, x}} - 1 \right).
\end{align}
Note that all integrals giving solutions of this kind can be traced back
to the common graph in Fig.~\ref{fig:BT3}.  We refrain from giving any
explicit results here, instead we refer to
Refs.~\cite{Anzai:2015wma,progdata}.

\section{Conclusion}

We described the computation of a contribution to the partonic cross
section for Higgs boson production via gluon fusion to N$^3$LO.  That
is, the sub-process initiated by two quarks of different flavour.  We
obtained analytic results with exact dependence on the Higgs boson mass
and the partonic center-of-mass energy.  New types of iterated integrals
beyond Harmonic Polylogarithms appear in the final expression.

This work represents an important step towards an exact result for all
N$^3$LO contributions to Higgs boson production.  To date only an
expansion around the soft limit is available from the
literature~\cite{Anastasiou:2014vaa,Anastasiou:2014lda,Li:2014afw,Anastasiou:2015ema}
which is, however, sufficient for phenomenology.  Performing an
expansion of our result around the threshold, we find for leading
logarithms agreement with the results from
Ref.~\cite{Anastasiou:2014lda}.

\section{Acknowledgments}

We would like to thank our collaborators C.~Anzai, A.~Hasselhuhn,
M.~H\"oschele, W.~Kilgore, M.~Steinhauser and T.~Ueda.  Parts of this
work were supported by the European Commission through contract
PITN-GA-2012-316704~(HIGGSTOOLS).

% -- bibliography ------------------------------------------------------


\begin{thebibliography}{999}

%1
\bibitem{Aad:2012tfa}
  G.~Aad {\it et al.} [ATLAS Collaboration],
  Phys.\ Lett.\ B {\bf 716}, 1 (2012)
  [arXiv:1207.7214 [hep-ex]].

%2
\bibitem{Chatrchyan:2012xdj}
  S.~Chatrchyan {\it et al.} [CMS Collaboration],
  Phys.\ Lett.\ B {\bf 716}, 30 (2012)
  [arXiv:1207.7235 [hep-ex]].

%3
\bibitem{Chetyrkin:1997un}
  K.~G.~Chetyrkin, B.~A.~Kniehl and M.~Steinhauser,
  Nucl.\ Phys.\ B {\bf 510} (1998) 61
  [hep-ph/9708255].

%4
\bibitem{Schroder:2005hy}
  Y.~Schr\"oder and M.~Steinhauser,
  JHEP {\bf 0601} (2006) 051,
  arXiv:hep-ph/0512058.

%5
\bibitem{Chetyrkin:2005ia}
  K.~G.~Chetyrkin, J.~H.~K\"uhn and C.~Sturm,
  Nucl.\ Phys.\  B {\bf 744} (2006) 121,
  arXiv:hep-ph/0512060.

%6
\bibitem{Spiridonov:1984br}
  V.~P.~Spiridonov,
  IYaI-P-0378.

%7
\bibitem{Tarasov:1980au}
  O.~V.~Tarasov, A.~A.~Vladimirov and A.~Y.~Zharkov,
  Phys.\ Lett.\ B {\bf 93} (1980) 429.

%8
\bibitem{Larin:1993tp}
  S.~A.~Larin and J.~A.~M.~Vermaseren,
  Phys.\ Lett.\ B {\bf 303} (1993) 334
  [hep-ph/9302208].

%9
\bibitem{Anastasiou:2002yz}
  C.~Anastasiou and K.~Melnikov,
  Nucl.\ Phys.\ B {\bf 646}, 220 (2002)
  [hep-ph/0207004].

%10
\bibitem{Wilczek:1977zn}
  F.~Wilczek,
  Phys.\ Rev.\ Lett.\  {\bf 39} (1977) 1304.

%11
\bibitem{Ellis:1979jy}
  J.~R.~Ellis, M.~K.~Gaillard, D.~V.~Nanopoulos and C.~T.~Sachrajda,
  Phys.\ Lett.\  B {\bf 83} (1979) 339.

%12
\bibitem{Georgi:1977gs}
  H.~M.~Georgi, S.~L.~Glashow, M.~E.~Machacek and D.~V.~Nanopoulos,
  Phys.\ Rev.\ Lett.\  {\bf 40} (1978) 692.

%13
\bibitem{Rizzo:1979mf}
  T.~G.~Rizzo,
  Phys.\ Rev.\  D {\bf 22} (1980) 178
  [Addendum-ibid.\  D {\bf 22} (1980) 1824].

%14
\bibitem{Dawson:1990zj}
  S.~Dawson,
  Nucl.\ Phys.\  B {\bf 359} (1991) 283.

%15
\bibitem{Spira:1995rr}
  M.~Spira, A.~Djouadi, D.~Graudenz and P.~M.~Zerwas,
  Nucl.\ Phys.\  B {\bf 453} (1995) 17,
  arXiv:hep-ph/9504378.

%16
\bibitem{Harlander:2002wh}
  R.~V.~Harlander and W.~B.~Kilgore,
  Phys.\ Rev.\ Lett.\  {\bf 88} (2002) 201801,
  arXiv:hep-ph/0201206.

%17
\bibitem{Ravindran:2003um}
  V.~Ravindran, J.~Smith and W.~L.~van Neerven,
  Nucl.\ Phys.\  B {\bf 665} (2003) 325,
  arXiv:hep-ph/0302135.

%18
\bibitem{Harlander:2009bw}
  R.~V.~Harlander and K.~J.~Ozeren,
  Phys.\ Lett.\  B {\bf 679} (2009) 467
  [arXiv:0907.2997 [hep-ph]].

%19
\bibitem{Pak:2009bx}
  A.~Pak, M.~Rogal and M.~Steinhauser,
  Phys.\ Lett.\ B {\bf 679} (2009) 473
  [arXiv:0907.2998 [hep-ph]].

%20
\bibitem{Harlander:2009mq}
  R.~V.~Harlander and K.~J.~Ozeren,
  JHEP {\bf 0911} (2009) 088
  [arXiv:0909.3420 [hep-ph]].

%21
\bibitem{Pak:2009dg}
  A.~Pak, M.~Rogal and M.~Steinhauser,
  JHEP {\bf 1002}, 025 (2010)
  [arXiv:0911.4662 [hep-ph]].

%22
\bibitem{Harlander:2009my}
  R.~V.~Harlander, H.~Mantler, S.~Marzani and K.~J.~Ozeren,
  Eur.\ Phys.\ J.\ C {\bf 66} (2010) 359
  [arXiv:0912.2104 [hep-ph]].

%23
\bibitem{Pak:2011hs}
  A.~Pak, M.~Rogal and M.~Steinhauser,
  JHEP {\bf 1109} (2011) 088
  [arXiv:1107.3391 [hep-ph]].

%24
\bibitem{Moch:2004pa}
  S.~Moch, J.~A.~M.~Vermaseren and A.~Vogt,
  Nucl.\ Phys.\ B {\bf 688} (2004) 101
  [hep-ph/0403192].

%25
\bibitem{Vogt:2004mw}
  A.~Vogt, S.~Moch and J.~A.~M.~Vermaseren,
  Nucl.\ Phys.\ B {\bf 691} (2004) 129
  [hep-ph/0404111].

%26
\bibitem{Anastasiou:2012kq}
  C.~Anastasiou, S.~Buehler, C.~Duhr and F.~Herzog,
  JHEP {\bf 1211} (2012) 062
  [arXiv:1208.3130 [hep-ph]].

%27
\bibitem{Hoschele:2012xc}
  M.~H\"oschele, J.~Hoff, A.~Pak, M.~Steinhauser and T.~Ueda,
  Phys.\ Lett.\ B {\bf 721} (2013) 244
  [arXiv:1211.6559 [hep-ph]].

%28
\bibitem{Buehler:2013fha}
  S.~Buehler and A.~Lazopoulos,
  JHEP {\bf 1310} (2013) 096
  [arXiv:1306.2223 [hep-ph]].

%29
\bibitem{Hoeschele:2013gga}
  M.~H\"oschele, J.~Hoff, A.~Pak, M.~Steinhauser and T.~Ueda,
  Comput.\ Phys.\ Commun.\  {\bf 185} (2014) 528
  [arXiv:1307.6925].

%30
\bibitem{Baikov:2009bg}
  P.~A.~Baikov, K.~G.~Chetyrkin, A.~V.~Smirnov, V.~A.~Smirnov and
  M.~Steinhauser,
  Phys.\ Rev.\ Lett.\  {\bf 102} (2009) 212002
  [arXiv:0902.3519 [hep-ph]].

%31
\bibitem{Gehrmann:2010ue}
  T.~Gehrmann, E.~W.~N.~Glover, T.~Huber, N.~Ikizlerli and C.~Studerus,
  JHEP {\bf 1006} (2010) 094
  [arXiv:1004.3653 [hep-ph]].

%32
\bibitem{Anastasiou:2013mca}
  C.~Anastasiou, C.~Duhr, F.~Dulat, F.~Herzog and B.~Mistlberger,
  JHEP {\bf 1312} (2013) 088
  [arXiv:1311.1425 [hep-ph]].

%33
\bibitem{Kilgore:2013gba}
  W.~B.~Kilgore,
  Phys.\ Rev.\ D {\bf 89} (2014) 073008
  [arXiv:1312.1296 [hep-ph]].

%34
\bibitem{Dulat:2014mda}
  F.~Dulat and B.~Mistlberger,
  arXiv:1411.3586 [hep-ph].

%35
\bibitem{Duhr:2014nda}
  C.~Duhr, T.~Gehrmann and M.~Jaquier,
  arXiv:1411.3587 [hep-ph].

%36
\bibitem{Anastasiou:2014vaa}
  C.~Anastasiou, C.~Duhr, F.~Dulat, E.~Furlan, T.~Gehrmann, F.~Herzog and
  B.~Mistlberger,
  Phys.\ Lett.\ B {\bf 737} (2014) 325
  [arXiv:1403.4616 [hep-ph]].

%37
\bibitem{Li:2014afw}
  Y.~Li, A.~von Manteuffel, R.~M.~Schabinger and H.~X.~Zhu,
  arXiv:1412.2771 [hep-ph].

%38
\bibitem{Anastasiou:2015ema}
  C.~Anastasiou, C.~Duhr, F.~Dulat, F.~Herzog and B.~Mistlberger,
  arXiv:1503.06056 [hep-ph].

%39
\bibitem{Anastasiou:2013srw}
  C.~Anastasiou, C.~Duhr, F.~Dulat and B.~Mistlberger,
  JHEP {\bf 1307} (2013) 003
  [arXiv:1302.4379 [hep-ph]].

%40
\bibitem{Hoschele:2014qsa}
  M.~H\"oschele, J.~Hoff and T.~Ueda,
  JHEP {\bf 1409} (2014) 116
  [arXiv:1407.4049 [hep-ph]].

%41
\bibitem{Anzai:2015wma}
  C.~Anzai, A.~Hasselhuhn, M.~H\"oschele, J.~Hoff, W.~Kilgore, M.~Steinhauser and T.~Ueda,
  JHEP {\bf 1507}, 140 (2015)
  [arXiv:1506.02674 [hep-ph]].

%42
\bibitem{Henn:2013pwa}
  J.~M.~Henn,
  Phys.\ Rev.\ Lett.\  {\bf 110} (2013) 25,  251601
  [arXiv:1304.1806 [hep-th]].

%43
\bibitem{Nogueira:1991ex}
  P.~Nogueira,
  J.\ Comput.\ Phys.\  {\bf 105} (1993) 279.

%44
\bibitem{diss_Hoff}
  J. Hoff, ``Methods for multiloop calculations and Higgs boson production at
  the LHC'', Dissertation, KIT, 2015.

%45
\bibitem{Harlander:1997zb}
  R.~Harlander, T.~Seidensticker and M.~Steinhauser,
  Phys.\ Lett.\ B {\bf 426} (1998) 125
  [hep-ph/9712228].

%46
\bibitem{Seidensticker:1999bb}
  T.~Seidensticker,
  hep-ph/9905298.

%47
\bibitem{Vermaseren:2000nd}
  J.~A.~M.~Vermaseren,
  math-ph/0010025.

%48
\bibitem{Kuipers:2012rf}
  J.~Kuipers, T.~Ueda, J.~A.~M.~Vermaseren and J.~Vollinga,
  Comput.\ Phys.\ Commun.\  {\bf 184} (2013) 1453
  [arXiv:1203.6543 [cs.SC]].

%49
\bibitem{Smirnov:2014hma}
  A.~V.~Smirnov,
  Comput.\ Phys.\ Commun.\  {\bf 189} (2014) 182
  [arXiv:1408.2372 [hep-ph]].

%50
\bibitem{Grigo:2014oqa}
  J.~Grigo and J.~Hoff,
  PoS LL {\bf 2014} (2014) 030
  [arXiv:1407.1617 [hep-ph]].

%51
\bibitem{progdata}
{\tt http://www.ttp.kit.edu/Progdata/ttp15/ttp15-019/}

%52
\bibitem{Anastasiou:2014lda}
  C.~Anastasiou, C.~Duhr, F.~Dulat, E.~Furlan, T.~Gehrmann, F.~Herzog and
  B.~Mistlberger,
  arXiv:1411.3584 [hep-ph].

\end{thebibliography}
\end{document}